\documentclass[11pt,a4paper]{article}
\usepackage[margin=2.3cm]{geometry}
\usepackage[T1]{fontenc}
\usepackage[utf8]{inputenc}
\usepackage{lmodern}
\usepackage{amsmath,amssymb}
\usepackage{microtype}
\usepackage{todonotes}
\usepackage{parskip}
\usepackage{enumerate}
\usepackage{tikz}
\usetikzlibrary{arrows.meta,positioning,calc}
\usepackage[font=small,labelfont=bf]{caption}
\usepackage{fancyhdr}
\usepackage[hidelinks]{hyperref}
\hypersetup{
  pdftitle={Algebraic Quantum Field Theory},
  pdfsubject={Chapter draft for Comprehensive Quantum Physics},
  pdfauthor={Kasia Rejzner},
  pdfkeywords={AQFT, local observables, quantum field theory}
}

\definecolor{before}{RGB}{45,115,142}
\definecolor{after}{RGB}{143,72,110}
\definecolor{interaction}{RGB}{205,136,57}

\usetikzlibrary{arrows.meta}

\newcommand{\Acal}{\mathcal{A}}
\newcommand{\Hcal}{\mathcal{H}}
\newcommand{\Wcal}{\mathcal{W}}

\newcommand{\RR}{\mathbb{R}}
\newcommand{\NN}{\mathbb{N}}
\newcommand{\MM}{\mathbb{M}}
\newcommand{\Bf}{\mathfrak{B}}
\newcommand{\Af}{\mathfrak{A}}

\newtheorem{theorem}{Theorem}

\newtheorem{definition}[theorem]{Definition}

\newtheorem{exa}[theorem]{Example}

\definecolor{retarded}{RGB}{0,100,156}
\definecolor{advanced}{RGB}{171,74,23}
\begin{document}

\title{Algebraic Quantum Field Theory}
\author{Kasia Rejzner}

\maketitle
\section*{Abstract}

Algebraic quantum field theory describes a quantum system through its observables, their localisation in spacetime, and the states that assign them expectation values. The separation of the specification of observables from the choice of a state makes it possible to compare vacuum, thermal and charged representations without assuming that they share a preferred Hilbert space. We develop the framework from familiar quantum mechanics, construct the free scalar field, and explain how causality and positive energy constrain local measurements. The Reeh--Schlieder theorem, the structure of local von Neumann algebras and the split property reveal key differences between quantum theory in the continuum (i.e. QFT) and ordinary quantum mechanics with finitely many degrees of freedom. Superselection theory relates observable charges to statistics and internal symmetries. Finally, local covariance and perturbative and homological methods extend the discussion to curved spacetime, interactions and gauge theories. We also discuss measurement theory and quantum information aspects of AQFT.


\section{The algebraic approach to quantum theory}\label{sec:1}

A quantum field theory must tell us which measurements can be made in a laboratory, how their outcomes are correlated, and how those correlations depend on the preparation of the system. These questions suggest separating two kinds of objects. The observables and their algebraic relations describe the system. A state supplies the expectation values that describe its preparation. Localisation supplies a third ingredient: every measurement belongs to a region of spacetime. This is important when we want to take special relativity effects into account and the notion of \emph{causality} plays a crucial role.

In elementary quantum mechanics the separation between observables and states can be easy to miss. One usually begins with a Hilbert space, introduces operators, and represents a preparation by a vector or a density matrix. This works well for a finite collection of canonical degrees of freedom. For infinitely many degrees of freedom, however, physically relevant representations of the same observable relations need not be unitarily equivalent. Moreover, a generic curved spacetime has no time translation symmetry with which to select a distinguished representation (called the \emph{vacuum}). An account of the observables should survive these changes of representation and background.

The Haag--Kastler approach implements this idea by assigning an algebra to each spacetime region \cite{HK}. Let us first specify the appropriate class of spacetimes. We will be working with Lorentzian manifolds, typically in 4 dimensions, where the metric has signature $(+---)$ (one often also uses the convention $(-+++)$). Among all Lorentzian manifolds, we distinguish the \emph{globally hyperbolic} ones. They don't have closed timelike curves and admit foliation into Cauchy surfaces.

In a time-oriented Lorentzian spacetime \(M\), a Cauchy surface is a subset \(\Sigma\subset M\) that every inextendible timelike curve intersects exactly once. Here, inextendible means that the curve cannot be continued further into either the past or the future within \(M\). Cauchy surfaces are important for specifying the dynamics of a given theory. Classically, if \emph{initial data} specified on such a hypersurface uniquely determines solutions of a given differential equation, we say that the Cauchy problem (i.e. dynamics) is well posed. 

A primary example of a globally hyperbolic spacetime is $n$-dimensional \emph{Minkowski spacetime} $\MM$, which is $\RR^{n}$ equipped with the diagonal metric with $1$ in the top left corner and $-1$ in the remaining diagonal entries.

Following \cite{HK} and the later generalizations to curved spacetimes \cite{BFV} we assign to every relatively compact, causally convex (contains all the causal curves starting and ending within it) open region $O$ the algebra $\Acal(O)$ of observables that can be measured in that region.

\begin{equation}\label{eq:1}
O\longmapsto\mathcal A(O)\,.
\end{equation}

On Minkowski spacetime, relatively compact regions are simply the bounded regions, and a good class of causally convex bounded regions is provided by the 
\emph{double cones}. A double cone is the intersection of the chronological future of one point with the chronological past of a later point. Physically, we can think of it as a finite laboratory with finite duration.

Results of measurements of such observables can be described after specifying \emph{a state}. Here a state is understood a complex-linear functional satisfying
\begin{equation}\label{eq:3}
	\omega(\mathbf1)=1,\qquad \omega(A^*A)\geq0.
\end{equation}
A Hilbert space representation can then be recovered from the state using the \emph{GNS construction} described in Section~\ref{sec:GNS}.

A state is mixed if it admits a nontrivial convex decomposition
\[
\omega=\lambda\omega_1+(1-\lambda)\omega_2,
\qquad 0<\lambda<1,
\]
where \(\omega_1\) and \(\omega_2\) are distinct states. Otherwise, it is pure. Thus, mixed states are precisely the states that are not extreme points of the convex state space.

This view on QFT where observables are primary objects and states are secondary is called \emph{algebraic quantum field theory} (AQFT). It studies both general consequences of this structure and the construction of theories that realise it. The most significant structural results in AQFT are:
\begin{itemize}
		\item \textbf{Superselection sectors and gauge symmetry.}
		DHR (Doplicher-Haag-Roberts) theory derives charge composition, conjugation and statistics from
		localized representations of the observable algebra. Under suitable
		assumptions, Doplicher--Roberts reconstruction also recovers a field algebra and a compact internal symmetry group 
		\cite{DHR1971,DHR1974,DR1990}.
		
		\item \textbf{Modular theory and spacetime geometry.}
		The Bisognano--Wichmann theorem identifies vacuum modular evolution for
		wedge algebras with Lorentz boosts, establishing a connection between
		localization, spacetime symmetry and the thermal character of the
		vacuum for accelerated observers \cite{BW1976}.
		
		\item \textbf{Spin--statistics and CPT.}
		Algebraic theorems establish the relation between spin and statistics,
		and the existence of CPT (charge, parity and time reversal) symmetry \cite{GuidoLongo1995}.
		
		\item \textbf{Reeh--Schlieder theorem.}
		Under standard vacuum, spectrum and additivity assumptions, this theorem expresses the extensive correlations of the vacuum \cite{ReehSchlieder1961}, see Section~\ref{sec:RS}.
		
		\item \textbf{Universal structure of local algebras.}
		Suitable scaling and phase-space assumptions imply that local algebras
		are hyperfinite type \(\mathrm{III}_1\) factors. This has consequences for entanglement entropy calculations
		\cite{Fredenhagen1985,BDF1987}.
		
		\item \textbf{Nuclearity and the split property.}
		Appropriate bounds on local degrees of freedom imply the existence of
		an intermediate type~I factor between suitably nested local algebras,
		yielding statistical independence and normal product states for
		separated regions \cite{BuchholzWichmann1986}.
		
		\item \textbf{Scattering theory.}
		Haag--Ruelle theory constructs asymptotic multiparticle states from
		local observables under suitable spectral assumptions for stable
		massive particles. Asymptotic completeness requires additional,
		model-dependent analysis \cite{Haag1958,Ruelle1962}.
		
		\item \textbf{Infraparticles and Gauss's law.}
		Under appropriate assumptions on asymptotic electric flux, electrically
		charged states cannot have sharp mass, identifying a structural
		obstruction to the ordinary Wigner particle description in charged
		sectors \cite{Buchholz1986}.
		
		\item \textbf{Equilibrium and complete passivity.}
		Completely passive states (those permitting no work extraction by
		cyclic processes from any number of independent copies) are precisely
		ground states or KMS states, giving an operational characterization of
		thermal equilibrium \cite{PuszWoronowicz1978}.
		
\end{itemize}

More recently, AQFT has also been very successful in describing QFT on curved spacetimes (Section~\ref{sec:8}), renormalisation and perturbative QFT (Section~\ref{sec:9}) as well as measurement in a relativistic setting (Section~\ref{sec:FV}).

In this chapter, we review the basics of AQFT and briefly cover the current directions of research.

We assume familiarity with quantum mechanics, special relativity and the free scalar field. Unless stated otherwise, the spacetime dimension is four, the metric has signature \((+---)\), and \(c=\hbar=1\). We restore \(\hbar\) when discussing deformation quantisation.

\section{Observables and states}\label{sec:2}

\subsection{The algebraic ingredients}

A complex unital star algebra has addition, multiplication, a unit and an operation \(A\mapsto A^*\), called \emph{involution}, satisfying
\begin{equation}\label{eq:2}
(AB)^*=B^*A^*,\qquad
(\lambda A+\mu B)^*=\overline{\lambda} A^*+\overline{\mu} B^*,\qquad
(A^*)^*=A.
\end{equation}

These are the familiar rules for the adjoint of an operator. An algebra equipped with such operation is called \emph{a star (or involutive) algebra}. Multiplication need not commute. In the bounded-observable formulation below, self-adjoint elements represent observables; positive elements between zero and the unit represent measurement effects. A projection is a special effect associated with an ideal yes-or-no test.

A topological algebra is an algebra equipped with a topology in such a way that all the algebra operations are continuous maps. If the topology comes from a norm $\|.\|$, the space is complete under this topology, and the norm satisfies $\|AB\|\leq \|A\|\|B\|$, then we have a \emph{Banach algebra}.

In a \(C^*\)-algebra the norm additionally satisfies \(\|A^*A\|=\|A\|^2\). The norm controls the size of an observable independently of a chosen state. The algebra of all bounded operators on a Hilbert space is the standard example of a \(C^*\)-algebra, but the definition does not presuppose a particular underlying Hilbert space. Polynomial algebras of smeared fields are usually topological star algebras of unbounded quantities; exponentiated fields lead instead to bounded generators and a \(C^*\)-algebra.

Let $\Acal$ be a $C^*$-algebra of observables and $\omega$ a state on $\Acal$. For \(A\in \Acal\), the number \(\omega(A)\) is its expectation value. Positivity gives probabilities between zero and one for effects. It also implies a Cauchy--Schwarz inequality,
\begin{equation}\label{eq:4}
|\omega(A^*B)|^2\leq\omega(A^*A)\,\omega(B^*B).
\end{equation}

One can derive this by expanding \(\omega((A+zB)^*(A+zB))\geq0\) and choosing the complex number \(z\) to minimise the result. Thus positivity already contains much of the geometry normally introduced through inner products.

States form a convex set. If a preparation randomly chooses between two states, its expectation values are their convex combination. A pure state is an extreme point of this set: it cannot be written as a nontrivial probabilistic mixture of different states. 

\subsection{GNS construction}\label{sec:GNS}

The Gelfand--Naimark--Segal construction turns a state into a Hilbert space representation. Its central idea is elementary: use possible algebraic operations (i.e. observable algebra elements) as candidate vectors, and identify operations that have zero norm.

For a state $\omega$ on a unital \(C^*\)-algebra, set
\begin{equation}\label{eq:5}
\mathcal N_\omega=\{A:\omega(A^*A)=0\},\qquad
\langle[A],[B]\rangle=\omega(A^*B).
\end{equation}

Here \([A]\) is the equivalence class of \(A\) in \(\mathcal A/\mathcal N_\omega\). Cauchy--Schwarz makes the inner product well defined. The set \(\mathcal N_\omega\) is a left ideal, so multiplication from the left gives an operator
\begin{equation}\label{eq:6}
\pi_\omega(C)[A]=[CA],\qquad \Omega_\omega=[\mathbf1].
\end{equation}

The inequality \(\omega(A^*C^*CA)\leq\|C\|^2\omega(A^*A)\) shows that this operator is bounded. Completing the quotient gives a Hilbert space \(\mathcal H_\omega\), and
\begin{equation}\label{eq:7}
\omega(A)=\langle\Omega_\omega,\pi_\omega(A)\Omega_\omega\rangle,
\qquad
\overline{\pi_\omega(\mathcal A)\Omega_\omega}=\mathcal H_\omega.
\end{equation}

The second condition says that \(\Omega_\omega\) is cyclic, i.e. every vector can be approximated by applying represented algebra elements to it.
Thus we have established how to obtain the triple $(\Hcal_\omega,\pi_\omega, \Omega_\omega)$ from a state $\omega$. It is easy to see that any such triple of a Hilbert space, representation and distinguished state vector yields a state by means of the first equation in \eqref{eq:7}.

 The cyclic representation $(\Hcal_\omega,\pi_\omega, \Omega_\omega)$ for a given $\omega$ is unique up to a unitary that intertwines the representations and identifies their distinguished vectors. For general star algebras there is an analogous construction on a common dense domain.

\begin{exa}
Take \(\mathcal A=M_n(\mathbb C)\) and \(\omega(A)=\operatorname{Tr}(\rho A)\), where \(\rho\) is a strictly positive density matrix. The GNS Hilbert space can be identified with the matrices equipped with the Hilbert--Schmidt inner product, i.e. for \(A,B\in M_{n}(\mathbb C)\),
\[
\langle A,B\rangle_{\mathrm{HS}}
=\operatorname{Tr}(A^\dagger B)
=\sum_{i=1}^{n}\sum_{j=1}^{n}\overline{A_{ij}}\,B_{ij},
\]
where \(A^\dagger\) denotes the conjugate transpose. 

The identification sends \([A]\) to \(A\rho^{1/2}\), and
\begin{equation}\label{eq:8}
\pi_\omega(C)X=CX,\qquad \Omega_\omega=\rho^{1/2},\qquad
\langle\rho^{1/2},C\rho^{1/2}\rangle=\operatorname{Tr}(\rho C).
\end{equation}

This space has dimension \(n^2\). Right multiplication commutes with every represented observable, so the representation is reducible. The distinguished vector therefore represents a mixed algebraic state. By contrast, if \(\rho=|\psi\rangle\langle\psi|\), the quotient removes a null subspace and the GNS representation reduces to the defining representation on \(\mathbb C^n\). In general a state on a \(C^*\)-algebra is pure exactly when its GNS representation is irreducible. 

\end{exa}

\section{Locality and relativistic dynamics}\label{sec:3}

\subsection{A net of local algebras}
Let $O$ range over all causally convex bounded regions of Minkowski spacetime $\MM$. As in \eqref{eq:1}, we assign $C^*$-algebras to such regions, requiring the following properties (axioms) to hold \cite{HK}:

\textbf{Isotony.} A larger laboratory contains all observables available in a smaller one:
\begin{equation}\label{eq:10}
O_1\subset O_2\quad\Longrightarrow\quad
\mathcal A(O_1)\subset\mathcal A(O_2).
\end{equation}

\textbf{Einstein causality.} If no causal curve connects the two regions, then the observables localised in these regions should be independent, so
\begin{equation}\label{eq:11}
[A,B]=0,\qquad A\in\mathcal A(O_1),\ B\in\mathcal A(O_2).
\end{equation}
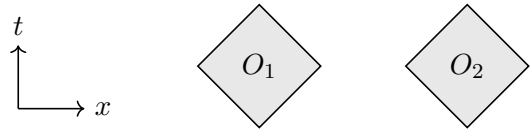
\begin{center}
	\begin{tikzpicture}[x=1.25cm,y=1.25cm,line width=0.6pt]
		\path[use as bounding box] (-2.2,-1) rectangle (2.3,1);
		\filldraw[fill=black!9] (-1.75,0)--(-1.1,.65)--(-.45,0)--(-1.1,-.65)--cycle;
		\filldraw[fill=black!9] (.45,0)--(1.1,.65)--(1.75,0)--(1.1,-.65)--cycle;
		\node at (-1.1,0) {$O_1$};
		\node at (1.1,0) {$O_2$};
		\draw[->] (1.35-5,-1.45+1)--(2.05-5,-1.45+1) node[right] {$x$};
		\draw[->] (1.35-5,-1.45+1)--(1.35-5,-.78+1) node[above] {$t$};
	\end{tikzpicture}
\end{center}
This is a condition on observable algebras. Fermionic field algebras use graded commutation; their observable subalgebras obey ordinary spacelike commutativity.

\textbf{Covariance.} A proper orthochronous Poincar\'{e} transformation \(g\) acts by an automorphism \(\alpha_g\) such that \(\alpha_g(\mathcal A(O))=\mathcal A(gO)\) and \(\alpha_g\alpha_h=\alpha_{gh}\). At this stage covariance is an action on observables. Its unitary implementation and continuity are properties to be established or imposed in a representation.

\textbf{Time slice.} The algebra of a neighbourhood \(N\) of a Cauchy surface of a causally convex region \(D\) is equal to  the algebra of that whole region:
\begin{equation}\label{eq:12}
\mathcal A(N)=\mathcal A(D).
\end{equation}

The time-slice condition expresses dynamical completeness of initial data. 

Often one requires an additional useful property called \textbf{Additivity.} When a region is covered by smaller regions, its algebra is generated by theirs, with the appropriate closure. A related assumption, \emph{weak additivity}, says that translates of any nonempty local region generate the global represented algebra. We will invoke this assumption when discussing the Reeh--Schlieder theorem in Section~\ref{sec:RS}.

In a \(C^*\)-formulation, the norm closure of the union of the bounded-region algebras is the \emph{quasilocal algebra} \(\mathcal A(\MM)\). A state $\omega$ on this algebra restricts consistently to every laboratory. By applying the GNS construction described in Section~\ref{sec:GNS}, one can represent this algebra on the Hilbert space.

In addition to the listed axioms, on Minkowski spacetimes one also requires further axioms related to the existence and properties of the \emph{vacuum state}, i.e. the distinguished  Poincaré-invariant ground state. This concept, however, does not generalize well to curved spacetimes. More about the vacuum can be found in the following sections.
\begin{exa}
This example illustrates the absence of signaling for spacelike separation. 
Suppose a local operation in \(O_1\) has finitely many Kraus operators \(K_i\in\mathcal A(O_1)\) with \(\sum_i K_i^*K_i=\mathbf1\). If its outcome is not selected, the state changes to \(\omega'(A)=\sum_i\omega(K_i^*AK_i)\). For \(B\in\mathcal A(O_2)\) in a spacelike separated region, locality gives
\begin{equation}\label{eq:13}
\omega'(B)=\sum_i\omega(BK_i^*K_i)=\omega(B).
\end{equation}

This calculation makes the operational content of commutativity explicit for operations of the stated local form. It does not say that the two regions are uncorrelated. Selecting one outcome gives a conditional state proportional to \(\omega(K_i^*AK_i)\), whose distant expectations can differ. To make use of that conditional ensemble, the distant experimenter must learn the selected outcome through ordinary communication. Correlation and signalling are different questions.
\end{exa}

\section{Building the free scalar field}\label{sec:4}

\subsection{Fields and commutators}\label{sec:commutators}
For the construction of the quantum free scalar field, we need the following input from classical theory:
\begin{equation}\label{eq:14}
P_{KG}=\Box+m^2,\qquad E=E^--E^+,\qquad
E(f,g)=\int f(x)(Eg)(x)\,d^4x\,,
\end{equation}
where $P_{KG}$ is the Klein--Gordon operator, $E^{\pm}$ are its retarded/advanced Green functions respectively, and $E$ is the commutator function (sometimes also called the \emph{causal propagator}). $E^{\pm}$ solve the inhomogeneous equation with support respectively to the causal future and past of the source. Their difference solves the homogeneous equation and vanishes between causally disjoint supports. The sign in \eqref{eq:14} fixes our commutator convention throughout the chapter.

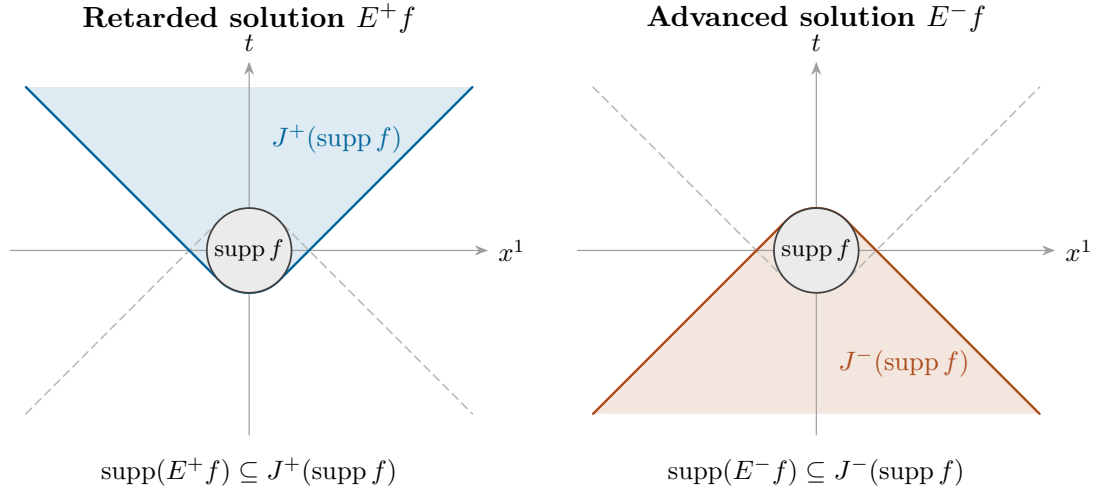
\begin{center}
	\begin{tikzpicture}[
		x=0.8cm,y=0.8cm,
		>=Stealth,
		line cap=round,line join=round,
		font=\small,
		axis/.style={->,gray!75,line width=0.45pt},
		inactive/.style={gray!55,densely dashed,line width=0.6pt},
		cone/.style={line width=0.9pt},
		source/.style={fill=gray!16,draw=black!75,line width=0.65pt}
		]
		
		\def\r{0.7}
		\def\T{2.7}
		\pgfmathsetmacro{\a}{\r/sqrt(2)}
		\pgfmathsetmacro{\outer}{\T+sqrt(2)*\r}
		
		\begin{scope}
			\node[font=\normalsize\bfseries] at (0,3.85)
			{Retarded solution $E^+f$};
			\fill[retarded!13]
			(-\outer,\T) -- (-\a,-\a)
			arc[start angle=225,end angle=315,radius=\r]
			-- (\outer,\T) -- cycle;
			\draw[inactive]
			(-\outer,-\T) -- (-\a,\a)
			arc[start angle=135,end angle=45,radius=\r]
			-- (\outer,-\T);
			\draw[axis] (-3.95,0) -- (3.95,0) node[right,text=black] {$x^1$};
			\draw[axis] (0,-3.05) -- (0,3.12) node[above,text=black] {$t$};
			\draw[cone,retarded]
			(-\outer,\T) -- (-\a,-\a)
			arc[start angle=225,end angle=315,radius=\r]
			-- (\outer,\T);
			\node[text=retarded] at (1.45,1.85)
			{$J^+(\operatorname{supp}f)$};
			\draw[source] (0,0) circle[radius=\r];
			\node[font=\footnotesize] at (0,0) {$\operatorname{supp}f$};
			\node at (0,-3.65)
			{$\operatorname{supp}(E^+f)\subseteq J^+(\operatorname{supp}f)$};
		\end{scope}
		
		\begin{scope}[xshift=7.5cm]
			\node[font=\normalsize\bfseries] at (0,3.85)
			{Advanced solution $E^-f$};
			\fill[advanced!14]
			(-\outer,-\T) -- (-\a,\a)
			arc[start angle=135,end angle=45,radius=\r]
			-- (\outer,-\T) -- cycle;
			\draw[inactive]
			(-\outer,\T) -- (-\a,-\a)
			arc[start angle=225,end angle=315,radius=\r]
			-- (\outer,\T);
			\draw[axis] (-3.95,0) -- (3.95,0) node[right,text=black] {$x^1$};
			\draw[axis] (0,-3.05) -- (0,3.12) node[above,text=black] {$t$};
			\draw[cone,advanced]
			(-\outer,-\T) -- (-\a,\a)
			arc[start angle=135,end angle=45,radius=\r]
			-- (\outer,-\T);
			\node[text=advanced] at (1.45,-1.85)
			{$J^-(\operatorname{supp}f)$};
			\draw[source] (0,0) circle[radius=\r];
			\node[font=\footnotesize] at (0,0) {$\operatorname{supp}f$};
			\node at (0,-3.65)
			{$\operatorname{supp}(E^-f)\subseteq J^-(\operatorname{supp}f)$};
		\end{scope}
%
		
	\end{tikzpicture}
\end{center}

A field is smeared with a smooth compactly supported test function: \(\Phi(f)=\int\Phi(x)f(x)\,d^4x\). Smearing describes a finite spacetime resolution and makes distributional expressions meaningful. For real test functions, the polynomial field algebra is generated by linear symbols satisfying
\begin{equation}\label{eq:15}
\Phi(P_{KG}f)=0,\qquad \Phi(f)^*=\Phi(f),\qquad
[\Phi(f),\Phi(g)]=iE(f,g)\mathbf1.
\end{equation}

Complex test functions are included by complex linearity. The first relation imposes the field equation, while the last is the covariant canonical commutation relation. Because \(P_{KG}\) is formally self-adjoint, \(E\) descends to the quotient of real test functions by \(P_{KG} C_c^\infty(M,\mathbb R)\), where the subscript $c$ indicates compact support. This quotient, with its nondegenerate antisymmetric form \(E\), is a real symplectic space \cite{Baer,DerGer13}.

The bounded version is the Weyl algebra, generated by unitaries \(\Wcal(f)\) labelled by the equivalence classes under the above quotient:
\begin{equation}\label{eq:16}
\Wcal(f)\Wcal(g)=e^{-iE(f,g)/2}\Wcal(f+g),\qquad
\Wcal(f)^*=\Wcal(-f),\qquad \Wcal(P_{KG}f)=\mathbf1.
\end{equation}

We suppress the brackets denoting equivalence classes. In a regular representation, \(t\mapsto\pi(\Wcal(tf))\) is strongly continuous and has a self-adjoint generator \(\Phi_\pi(f)\), so that \(\pi(\Wcal(tf))=e^{it\Phi_\pi(f)}\). 

Define \(\mathcal A(O)\) as the \(C^*\)-algebra generated by \(\Wcal(f)\) with \(\operatorname{supp}f\subset O\). Isotony follows from support inclusion. For spacelike separated supports, \(E(f,g)=0\), so the Weyl relations give commutativity. Thus the causal support of a classical differential equation becomes the locality of a quantum theory.

\subsection{Time-slice property}

The free field also illustrates the time-slice axiom. Let \(N\) be a causally convex region that contains a Cauchy surface. By using advanced and retarded solutions and a smooth cutoff that changes across \(N\), one can replace any test function \(f\) by a test function \(f_N\) supported in \(N\) such that
\begin{equation}\label{eq:17}
f-f_N=P_{KG}h\quad\text{for some }h\in C_c^\infty(M).
\end{equation}
An explicit construction has been given in \cite{CF09}.
Choose two Cauchy surfaces $\Sigma_-$ and $\Sigma_+$
such that the region between them is contained in $N$,
and a smooth function $\chi$ which vanishes in the
past of $\Sigma_-$, equals one in the future of
$\Sigma_+$, and satisfies $\operatorname{supp}(d\chi)
\subset N$. Define
\begin{equation}
	f_N=-P_{KG}(\chi Ef)
\end{equation}
Since $P_{KG}Ef=0$, $f_N$ vanishes in the future of $\Sigma_+$. Since $\chi=0$ in the past of $\Sigma_-$, $f_N$ vanishes also there. 
Together with global hyperbolicity and the fact that $f$ is compactly supported, we conclude that
$f_N\in C_c^\infty(N)$. Moreover, setting
\[
h=\chi E^-f+(1-\chi)E^+f\in C_c^\infty(M),
\]
we obtain
\[
P_{KG}h=f+P_{KG}(\chi Ef)=f-f_N.
\]
Hence $f-f_N=P_{KG}h$. Equation \eqref{eq:15} then gives \(\Phi(f)=\Phi(f_N)\), and the Weyl generators agree as well. 

\subsection{Gaussian states and the vacuum}\label{sec:Gaussian}

The relations do not yet specify expectation values. We need to specify a state. A convenient class of states is given by \emph{quasi-free} ones.

A quasifree, or centered Gaussian, state $\omega$ on the Weyl algebra is determined by a real symmetric bilinear form \(\mu\) obeying positivity and the uncertainty bound
\begin{equation}\label{eq:18}
\mu(f,f)\geq0,\qquad
|E(f,g)|^2\leq4\mu(f,f)\mu(g,g).
\end{equation}

On the symplectic quotient it is defined by
\begin{equation}\label{eq:19}
\omega(\Wcal(f))=e^{-\mu(f,f)/2}.
\end{equation}

Similarly, on the field algebra generated by $\Phi(f)$, a quasifree state $\omega$ is determined by its two-point function
\begin{equation}\label{eq:W}
\omega(\Phi(f)\Phi(g))\equiv 
W(f,g)=\mu(f,g)+\frac{i}{2}E(f,g)\,,
\end{equation}
where $\mu$ is again a real symmetric bilinear form. The uncertainty bound ensures positivity of the complex form $W$. Odd correlation functions vanish, while even ones are expressed in terms of the 2-point functions as sums over pairings.

In Minkowski spacetime the usual vacuum two-point function is a bi-distribution given by
\begin{equation}\label{eq:21}
W_0(x,y)=\int\frac{d^3\mathbf k}{(2\pi)^3\,2\omega_{\mathbf k}}
e^{-i\omega_{\mathbf k}(x^0-y^0)+i\mathbf k\cdot(\mathbf x-\mathbf y)},
\qquad \omega_{\mathbf k}=\sqrt{\mathbf k^2+m^2}.
\end{equation}

The integral is understood distributionally. Positivity follows after smearing because the result is an integral of a squared Fourier amplitude against a positive mass-shell measure. Its antisymmetric part is \(iE\), consistently with \eqref{eq:14}.

\begin{exa}
Now we illustrate local commutativity with nonzero correlation. Choose real \(f\) and \(g\) with spacelike separated supports. The commutator of $\Phi(f)$ and $\Phi(g)$  vanishes, but the two-point function in \eqref{eq:W} gives \(W(f,g)=\mu(f,g)\), which is generally nonzero. The absence of a causal response fixes the antisymmetric part of the two-point function. It does not force its symmetric part to vanish. This distinction is the simplest free-field example of the difference between causality and statistical independence.
\end{exa}

\section{Positive energy and the local structure of the vacuum}\label{sec:5}

\subsection{Selecting a vacuum representation}\label{sec:vacuum}

Suppose the net has a Poincar\'{e}-invariant state \(\omega_0\) on the quasilocal algebra $\Acal(\MM)$. In its GNS representation the Poincar\'{e} symmetry action
on $\Hcal_{\omega_0}$ is fixed by how it acts on the dense set of equivalence classes: \(U(g)[A]=[\alpha_g(A)]\). Invariance of the state makes this an isometry, which extends to a unitary and fixes the invariant cyclic vector \(\Omega_0\). Assuming the necessary continuity, translations have commuting self-adjoint generators $P_\mu$ with their joint spectrum $\operatorname{sp}(P)$ satisfying:
\begin{equation}\label{eq:22}
U(a)=e^{iP_\mu a^\mu},\qquad
\operatorname{sp}(P)\subset\overline V_+
=\{p:p^0\geq|\mathbf p|\}\,,
\end{equation}
where $\overline{V}_+$ denotes the closed future light cone in momentum space:
\[
\overline{V}_+
=\left\{p=(p^0,\mathbf p)\in\mathbb R^4:
p^0\geq|\mathbf p|\right\}.
\]
The joint spectrum is defined as follows. We start with the strongly commuting self-adjoint energy–momentum operators $P_\mu$, meaning that their spectral projections commute. They then possess a joint projection-valued measure $\mathsf E_P$ on $\mathbb R^4$, satisfying
\[
P_\mu=\int_{\mathbb R^4}p_\mu\,d\mathsf E_P(p).
\]
Their joint spectrum is the support of this measure: $\operatorname{sp}(P)=\operatorname{supp}\mathsf E_P$.

The spectrum condition expresses positive energy in every inertial frame. For the usual vacuum setting we also require that the invariant vector be unique up to scalar multiples. More precisely:
\begin{definition}
	A \emph{vacuum state} is a translationally invariant state obeying the spectrum condition,
	whose GNS vector is the unique translationally invariant vector (up to scalar multiples) in the GNS
	Hilbert space. The corresponding GNS representation is called the \emph{vacuum representation}.
\end{definition}

Positive energy has a consequence beyond stability. Matrix elements of translations admit analytic continuation into suitable complex spacetime directions. In physical terms, the sign of energy constrains how sharply correlation functions can vanish as functions of translation. This is the analytic input to the Reeh--Schlieder theorem.

\subsection{The Reeh--Schlieder theorem}\label{sec:RS}
We are now ready to state one of the key results in AQFT, \emph{the Reeh--Schlieder theorem}. We already know that in the GNS representation of $\Acal(\MM)$ for a given state $\omega$, the vector $\Omega_{\omega}$ is cyclic, i.e. any vector in $\Hcal_{\omega}$ can be approximated by acting with operators from $\Acal(\MM)$ on $\Omega_{\omega}$. We could play the same game with a state restricted to a local algebra $\Acal(O)$, i.e.
\[
\omega_O=\omega|_{\mathcal A(O)}.
\]
By GNS uniqueness, its representation is unitarily equivalent to the representation of $\mathcal A(O)$ on the subspace
\[
\mathcal H_{\omega,O}
=
\overline{\pi_\omega(\mathcal A(O))\Omega_\omega}
\subseteq\mathcal H_\omega.
\]
Remarkably, a much stronger result can be obtained, if the state in question is the vacuum. It turns out that for the vacuum state one can approximate \emph{every vector in $\Hcal_{\omega_0}$} by applying operations localised in \emph{any local region O}. This is the content of the Reeh--Schlieder theorem, which we will now explain in more detail.

For simplicity of notation we will now denote the GNS triple corresponding to the vacuum state $\omega_0$ by $\Hcal_0,\pi_0, \Omega_0$.

\begin{theorem}[Reeh--Schlieder]
Consider a net of $C^*$ algebras $O\mapsto \Acal(O)$, which in addition satisfies weak additivity in the vacuum state $\omega_0$ (as specified in Section~\ref{sec:vacuum}). Let $O$ be any causally convex bounded open region. Then:
\begin{enumerate}[(i)]
	\item vectors of the form $A\Omega_{0}$  for $A \in \pi_0(\Acal(O))$ are dense in $\Hcal_0$ (cyclic);
	\item if $A \in \pi_0(\Acal(O))$ annihilates the
	vacuum vector, $A\Omega_0 = 0$, then $A = 0$ (separating).
\end{enumerate}
\end{theorem}
A detailed proof can be found in \cite{FR2020}, based on \cite{Ara99}.

The proof of cyclicity begins with a vector orthogonal to all local excitations. Small translations of observables in a smaller double cone remain in \(O\), so suitable vacuum matrix elements vanish for an open set of translation parameters. Analytic continuation, supplied by positive energy, extends this vanishing. Weak additivity then enlarges the observables under consideration to the global algebra, whose cyclicity forces the orthogonal vector to be zero. This is an outline of the mechanism; the several-variable analytic continuation is the substantive theorem.

Separating follows from a shorter argument. If \(A\Omega_0=0\) and \(B\) lies in a spacelike separated algebra, then \(AB\Omega_0=BA\Omega_0=0\). The vectors \(B\Omega_0\) are dense by cyclicity, so the bounded operator \(A\) must vanish. One consequence is that a nonzero local projection cannot have exactly zero vacuum probability: \(\omega_0(Q)=\|Q\Omega_0\|^2>0\).

At first sight, cyclicity seems to allow an arbitrarily small laboratory to prepare any state in the universe. The missing ingredients in that reading are probability and resources. An approximating operator need not be unitary or have a uniformly bounded norm or energy cost. After rescaling it to an admissible measurement operator, the desired outcome may have a very small probability. The theorem is a density statement, not a protocol with a guaranteed outcome. The unselected local operation still obeys the calculation in \eqref{eq:13}.

The theorem expresses the extensive correlations of the vacuum, but it does not by itself establish every proposed entanglement measure or Bell inequality. Those questions require a choice of commuting algebras, states and measurement operations. The algebraic framework makes those choices explicit.

\subsection{Why local algebras are not ordinary tensor factors}

Given a represented algebra, its commutant (denoted with prime) consists of all bounded operators commuting with it. Its double commutant is its closure in the weak operator topology. For a local algebra $\mathcal A(O)$, we can represent it using a GNS representation $\pi_\omega$ coming from a state $\omega$. We define
\begin{equation}\label{eq:24}
\mathcal R(O)=\pi_\omega(\mathcal A(O))'',\qquad
\mathcal R(O)'=\{B:[B,A]=0\ \text{for all }A\in\mathcal R(O)\}.
\end{equation}
The weak closure allows limits defined through matrix elements. It depends on the representation, unlike the original abstract net. 

\begin{definition}
	A von Neumann algebra on a complex Hilbert space $\mathcal H$ is a unital $*$-subalgebra
	\[
	\mathcal M\subseteq\mathcal B(\mathcal H)
	\]
	that is closed in the weak operator topology.
\end{definition}
The center of a von Neumann algebra $\mathcal M$ is
\[
Z(\mathcal M)
=
\{A\in\mathcal M:AB=BA\text{ for every }B\in\mathcal M\}
=
\mathcal M\cap\mathcal M'.
\]
A von Neumann algebra is called a factor when its centre consists only of scalar multiples of the identity. This says that it has no nontrivial classical variable commuting with all its observables from within the algebra.

For ordinary bipartite quantum mechanics one writes \(\mathcal H=\mathcal H_A\otimes\mathcal H_B\) and represents the first subsystem by \(B(\mathcal H_A)\otimes\mathbf1\). This is a type I factor. It has minimal projections and the trace structure used to define a reduced density matrix.

Local algebras in familiar continuum relativistic theories have a different structure. Under suitable short-distance scaling and phase-space assumptions they are hyperfinite type \(\mathrm{III}_1\) factors \cite{yngvason2005role,Fredenhagen1985}. The assertion needs these additional hypotheses; it is not a consequence of the minimal net axioms alone. The type III property means that there is no nonzero normal semifinite trace and no nonzero finite projection in the operator-algebraic sense. In particular, there are no minimal projections and no pure normal states. The subscript 1 specifies a finer part of the classification, while hyperfiniteness describes approximation by finite-dimensional subalgebras in a weak topology.

Consequently, a sharp local region does not generally have an intrinsic reduced density matrix with a trace of the type used for \(B(\mathcal H_A)\). This does not mean that its expectation values cannot be represented by a density operator on an ambient Hilbert space. They can, for normal states. What fails is the identification of its local algebra with a full type I tensor factor and its associated intrinsic trace. Pure states on the underlying abstract algebra also exist; the restriction concerns normal states in the representation under discussion.

This distinction explains why a cutoff computation of spatial entanglement entropy needs care in the continuum limit. The expression \(-\operatorname{Tr}(\rho_O\log\rho_O)\) presupposes a subsystem density matrix. AQFT instead retains the local algebra and uses tools such as relative entropy and modular theory, which do not require a local trace. Relative entropy can itself be infinite for some pairs of states; the algebraic formulation does not promise finiteness without further conditions.

Even when individual local algebras share the same abstract factor type, different theories do not become identical. Their inclusion relations, relative positions, symmetry actions and distinguished states carry physical information. Specifying one factor without the net loses most of that information.

\subsection{The split property and laboratories with separation}\label{sec:split}

There is a controlled way to recover a tensor-product description. Let the closure of a double cone \(O_1\) lie inside a larger double cone \(O_2\). The split property requires a type I factor \(\mathcal N\) between their algebras (see e.g. \cite{fewster2016split} for a recent review):
\begin{equation}\label{eq:25}
\mathcal R(O_1)\subset\mathcal N\subset\mathcal R(O_2).
\end{equation}

The separation between the boundaries is essential. A unitary identification with a tensor product then makes \(\mathcal N\) act on one factor and \(\mathcal R(O_2)'\) on the other. Normal states in the inner region and in the exterior can be extended to a normal product state satisfying \(\omega(AB)=\omega_1(A)\omega_2(B)\) for \(A\in\mathcal R(O_1)\) and \(B\in\mathcal R(O_2)'\).

The type I algebra includes a choice of degrees of freedom in the intervening region. It is not the sharp algebra \(\mathcal R(O_1)\). Thus the split property is compatible with type III local algebras and with a correlated vacuum: it supplies independently preparable states, not a claim that the vacuum is a product state. Nor does splitting alone guarantee finite von Neumann entropy for every state on the intervening type I factor.

A physical route to splitting uses nuclearity conditions. These control the number of local excitations accessible at a given energy resolution. A typical map takes a bounded local operator \(A\) to \(e^{-\beta H}A\Omega_0\). Requiring appropriate nuclearity and growth bounds expresses that localisation does not produce too many low-energy degrees of freedom. Such conditions are additional restrictions on a theory, stronger than spacelike commutativity.

\section{Thermal equilibrium and modular structure}\label{sec:6}

A state at nonzero temperature is not characterised by the vacuum spectrum condition. In a finite system with Hamiltonian \(H\), the Gibbs state is \(\omega_\beta(A)=Z^{-1}\operatorname{Tr}(e^{-\beta H}A)\). In infinite volume, the trace and the partition function may not exist even when equilibrium correlation functions are meaningful.

The Kubo--Martin--Schwinger condition, or KMS condition, expresses equilibrium directly in terms of the state and a time evolution \(\alpha_t\). For suitable \(A,B\), there is a function analytic in \(0<\operatorname{Im}z<\beta\), continuous on its boundary, with
\begin{equation}\label{eq:26}
F_{A,B}(t)=\omega_\beta(A\alpha_t(B)),\qquad
F_{A,B}(t+i\beta)=\omega_\beta(\alpha_t(B)A).
\end{equation}

For a finite system, the condition follows from cyclicity of the trace. It remains meaningful for a \(C^*\) dynamical system or a suitably continuous von Neumann algebra evolution without a Gibbs density matrix \cite{HaagHugenholtzWinnink1967}. Equilibrium is always relative to the specified time flow.

\begin{exa}
In the Hilbert--Schmidt GNS realisation of a faithful Gibbs state, the implementing unitaries act as \(U(t)X=e^{itH}Xe^{-itH}\). Their generator is the Liouvillian \(L X=HX-XH\). If \(H|j\rangle=E_j|j\rangle\), then \(L(|j\rangle\langle k|)=(E_j-E_k)|j\rangle\langle k|\). Its spectrum generally has both signs, even when \(H\) is positive. Thus invariance of a thermal state under time translations does not make its canonical GNS implementation a positive-energy vacuum representation.
\end{exa}

Modular theory associates an intrinsic flow with a von Neumann algebra and a cyclic separating vector. Starting with the antilinear operator \(S_0A\Omega=A^*\Omega\), one closes it and takes the polar decomposition
\begin{equation}\label{eq:27}
S=J\Delta^{1/2},\qquad
\sigma_s(A)=\Delta^{is}A\Delta^{-is}.
\end{equation}

Here \(J\) is antiunitary and \(\Delta\) is a positive operator. Tomita--Takesaki theory states that \(\sigma_s\) preserves the algebra and that \(J\) maps it to its commutant. For the finite Gibbs example, \(\sigma_s(A)=\rho^{is}A\rho^{-is}=\alpha_{-\beta s}(A)\). This fixes the relation between modular parameter and the KMS convention in \eqref{eq:26}.

In the setting of the Bisognano--Wichmann theorem, the modular flow of the Minkowski vacuum restricted to a wedge is a Lorentz boost flow, with the parameter scaled by \(2\pi\). An observer with constant proper acceleration \(a\) follows a boost orbit, so the wedge vacuum is thermal with respect to that observer's proper time at
\begin{equation}\label{eq:28}
T_{\mathrm U}=\frac{\hbar a}{2\pi c k_{\mathrm B}}.
\end{equation}

This is the Unruh temperature. The theorem applies to appropriate relativistic field theories satisfying its hypotheses; geometric modular flow is not an automatic property of an arbitrary region in an arbitrary net. Wedges are a particularly informative case because their symmetry makes the relation between geometry and thermal structure explicit.

More generally, we say that the flow of the modular Hamiltonian 
 is \emph{geometric} if the modular automorphisms map subregion algebras
according to a spacetime flow preserving $O$. Such geometric action
is known only under additional hypotheses:
\begin{itemize}
	\item \textbf{Wedges.}
	The Bisognano--Wichmann theorem identifies vacuum modular flow
	with Lorentz boosts~\cite{BisognanoWichmann1975,BW1976}.
	Algebraic extensions cover suitable asymptotically complete massive
	and massless theories~\cite{Mund2001,DybalskiMorinelli2020};
	the property does not follow from the minimal Haag--Kastler axioms.
	
	\item \textbf{Conformal double cones.}
	In conformally covariant theories, vacuum modular flow for a double
	cone is the conformal flow preserving that
	region~\cite{HislopLongo1982,BrunettiGuidoLongo1993}.
	Explicit one-particle generators are available for the free massless
	scalar field~\cite{LongoMorsella2023}.
	
	\item \textbf{Special states in two-dimensional CFT.}
	Suitable states on disjoint intervals admit geometric modular
	action, although vacuum flow can mix different
	intervals~\cite{LongoMartinettiRehren2010}.
	Conformal transformations of vacuum and thermal states also realize
	broad classes of prescribed geometric flows satisfying appropriate
	boundary and regularity conditions~\cite{CaminitiEtAl2025}.
	
	\item \textbf{Curved spacetimes.}
	Geometric modular flow occurs for suitable equilibrium states
	associated with Killing horizons~\cite{Sewell1982}.
	Conformal transport yields further examples for conformal fields,
	including de Sitter diamonds~\cite{Frob2023}; these constructions
	depend on both the geometry and the chosen state.
	
	\item \textbf{Null surfaces.}
	For suitable regions bounded by cuts of a null plane, vacuum modular
	flow acts geometrically along the null generators, while its bulk
	action need not be geometric~\cite{CasiniTesteTorroba2017}.
	A rigorous operator-algebraic construction is available for scalar
	free fields~\cite{MorinelliTanimotoWegener2022}.
	
	\item \textbf{General restrictions.}
	Under suitable locality and nondegeneracy assumptions, smooth
	geometric modular flows must be
	conformal~\cite{Keyl1993,Sorce2024}.
	Analyticity and microlocal spectrum conditions provide further
	restrictions on their causal
	character~\cite{Sorce2024,PinamontiSandersVerch2019}.
	
	\item \textbf{Uniqueness of null-local flow.}
	For the massless Klein--Gordon field in Minkowski dimension
	$D\geq3$, a recent preprint proves that a faithful normal state on a
	diamond algebra in the vacuum representation, whose modular flow
	is an angle-preserving diffeomorphism of the past null boundary,
	must equal the vacuum restriction~\cite{SatishchandranSorce2026}.
	
	\item \textbf{Massive theories and general bounded regions.}
	No general geometric-flow theorem is available. Even for free
	fields, determining modular generators involves substantial
	numerical or perturbative analysis, as illustrated by massive
	scalar and fermionic
	models~\cite{BostelmannCadamuroMinz2023,CadamuroFrobMinz2025}.
	
	\item \textbf{Reconstruction from modular inclusions.}
	Half-sided modular inclusions reconstruct positive-energy
	translations and dilations; additional compatibility conditions
	yield further spacetime or conformal symmetries. This provides a
	structural route from modular data to
	geometry~\cite{Borchers1992,Wiesbrock1993,ArakiZsido2005}.
	
	\item \textbf{Instantaneously geometric modular flow (conjecture).}
	Jensen, Sorce and Speranza conjecture that suitably chosen states
	for general subregions admit modular Hamiltonians given by local
	stress-tensor integrals on a selected Cauchy surface, with
	boost-like behaviour near the entangling
	surface~\cite{JensenSorceSperanza2023}.
	The geometric interpretation concerns the infinitesimal action
	at that surface; geometric modular evolution throughout the
	region is not required. The general existence problem remains
	open.
\end{itemize}

\section{Charges and superselection sectors}\label{sec:7}
\subsection{DHR analysis and generalization}
The observable algebra can describe charges even when charged fields are not themselves observables. Suppose the Hilbert space decomposes into charge subspaces and all observables are block diagonal. Relative phases between vectors in different charge subspaces cannot be measured by that observable algebra. Its inequivalent irreducible representations then describe distinct superselection sectors.

The Doplicher--Haag--Roberts, or DHR, analysis makes localisation of a charge precise 		\cite{DHR1971,DHR1974,DR1990}. A representation \(\pi\) satisfies the DHR selection criterion when, outside any sufficiently chosen bounded localisation region, it is equivalent to the vacuum representation. In its standard transportable formulation this means that for every double cone \(O\),
\begin{equation}\label{eq:29}
\pi\big|_{\mathcal A(O')}\simeq
\pi_0\big|_{\mathcal A(O')}.
\end{equation}

Here \(O'\) is the spacelike complement and its algebra is generated by the bounded regions it contains. The criterion allows a representative of the charge to be localised in any double cone; it does not assert that one fixed charged state is indistinguishable from the vacuum outside every region simultaneously.

With assumptions including Haag duality in the vacuum representation, such sectors can be represented by transportable localised endomorphisms \(\rho\) of the observable algebra. Haag duality strengthens locality to \(\mathcal R(O)=\mathcal R(O')'\): the local algebra contains all bounded operators commuting with the exterior algebra. Localisation means \(\rho(A)=A\) for observables in the spacelike complement. The representation is equivalent to \(\pi_0\circ\rho\), and composition \(\rho\sigma\) describes combining charges. Intertwiners compare equivalent or related charge configurations. Moving localisation regions apart and exchanging them supplies statistics operators.

For the standard four-dimensional DHR setting with finite statistics and the required completeness properties, this structure leads to a compact internal gauge group and a field algebra whose invariant elements are the observables \cite{DR1990}. The reconstructed fields obey the corresponding Bose or Fermi commutation rules. What began as representation theory of observables recovers an internal symmetry and charged fields. The reconstructed compact group is a global internal symmetry of the field algebra; it should not be confused with the redundancy of an arbitrary local gauge description.

The localisation hypothesis also marks the theory's limits. In electrodynamics, Gauss's law ties electric charge to a long-range flux. Charged sectors therefore do not fit the simple bounded-region DHR criterion, and infrared effects require a broader treatment. 

Buchholz and Fredenhagen \cite{BuchholzFredenhagen1982} generalized the Doplicher--Haag--Roberts (DHR) analysis of superselection sectors by relaxing the requirement that charges be localizable in bounded spacetime regions. Instead, they considered representations whose restrictions to the observable algebras associated with the spacelike complements of unbounded, cone-like regions (spacelike cones) are unitarily equivalent to the vacuum representation. This weaker localization criterion accommodates charges that cannot be localized in double cones and allows for charged fields with string-like localization extending to spacelike infinity. \cite{BuchholzFredenhagen1982} proves that, under suitable spectral assumptions, massive single-particle states satisfy this localization criterion. In four spacetime dimensions, the resulting superselection structure retains essential features of DHR theory, including Bose--Fermi (para)statistics and the existence of conjugate charges corresponding to antiparticles.

\subsection{Different statistics}

The DHR analysis shows that the statistics of superselection sectors is determined by locality and the topology of spacetime \cite{DHR1971}. In spacetime dimensions \(d\geq 3\), sectors localized in bounded regions obey permutation-group statistics, allowing Bose, Fermi, and parastatistics. In \(1+1\) dimensions, however, the spacelike complement of a double cone has two disconnected components, leading to two inequivalent ways of ordering localized charges and potentially to braid-group statistics. In \(2+1\) dimensions, the Buchholz–Fredenhagen (BF) extension to charges localized in spacelike cones permits braid-group statistics as well, although ordinary DHR sectors localized in bounded regions retain permutation statistics \cite{BuchholzFredenhagen1982}.

These more general statistics are called exotic statistics. Unlike permutation statistics, where exchanging two identical particles twice acts trivially, braid statistics distinguishes different winding patterns of particle trajectories. In the simplest case, exchanges produce a phase \(e^{i\theta}\), giving rise to Abelian anyons, while more general sectors carry non-Abelian representations of the braid group, describing non-Abelian anyons (also called plektons). The corresponding superselection categories are braided rather than symmetric tensor categories and may have non-integer statistical dimensions. Furthermore, generalized spin-statistics theorems relate the statistics phase to the spin, which in \(2+1\) dimensions need not be integral or half-integral \cite{FredenhagenRehrenSchroer1989,FrohlichGabbiani1990,FredenhagenRehrenSchroer1992}. 

Recent experiments have provided striking realizations of non-Abelian anyons (plektons), whose exchange statistics is governed by noncommuting representations of the braid group. In particular, experiments using trapped-ion quantum processors have demonstrated non-Abelian braiding in engineered topologically ordered states \cite{Iqbal2024} and, more recently, the implementation of universal quantum gates through braiding and fusion of non-Abelian anyons \cite{Lo2026}. Kim and collaborators reported interference measurements in bilayer graphene at even-denominator fractional quantum Hall states \cite{Kim2026}. Their observations include changes in interference patterns consistent with localized quasiparticles of charge \(e/4\), as expected in candidate non-Abelian states. This is important because it concerns quasiparticles arising naturally in a condensed-matter system, rather than states deliberately prepared in a quantum processor. However, the experiment does not yet establish non-Abelian braid-group representations unambiguously.

Recent operator-algebraic work by Bols, Hamdan, Naaijkens and Vadnerkar \cite{Bols2026} proves, under appropriate assumptions, that the braided category of cone-localized sectors of Kitaev's quantum double model for a finite group \(G\) is equivalent to \(\operatorname{Rep}_f\mathcal D(G)\). 
These results provide a rigorous operator-algebraic description of non-Abelian anyonic sectors and demonstrate how the localization and superselection methods originating in AQFT can be applied to topologically ordered quantum many-body systems. Their relation to candidate non-Abelian phases in fractional quantum Hall systems is conceptual rather than an identification of the underlying topological orders.

\section{Quantum fields on curved spacetime}\label{sec:8}

\subsection{Local covariance}

On a generic curved spacetime there is no preferred global time translation, no conserved total energy and no canonical particle interpretation. Local hyperbolic equations and causal separation still make sense. This makes the algebraic separation of observables from states especially effective.

The principle of local covariance developed from earlier work on quantum field theory in curved spacetime, notably Dimock's generally covariant construction of local observable algebras \cite{Dimock1980} and Kay's principle of F-locality \cite{Kay1992}, which required consistency between the local algebras induced from a spacetime and those constructed intrinsically on globally hyperbolic subregions. Another motivation arose from the problem of perturbative renormalization on curved spacetimes. Building on microlocal methods and the Epstein–Glaser approach to renormalization \cite{EG}, Brunetti and Fredenhagen \cite{BF0} constructed perturbatively interacting quantum field theories on arbitrary globally hyperbolic spacetimes. Their construction, however, left unresolved the problem of fixing finite renormalizations in a manner depending locally and covariantly on the background geometry. 

The modern formulation emerged from discussions at an Oberwolfach workshop in September 2000, involving, among others, Brunetti, Fewster, Fredenhagen, Hollands, Kay and Verch (as acknowledged in a later Oberwolfach report \cite{Fredenhagen2013Oberwolfach}). Verch \cite{Verch2001} subsequently applied a locally generally covariant framework to the spin-statistics theorem, explicitly recognising its relationship with Kay's earlier work. Hollands and Wald \cite{HW} developed the principle as a condition on Wick polynomials and time-ordered products, acknowledging Fredenhagen's suggestion of the algebraic formulation during the Oberwolfach meeting. The general functorial framework for locally covariant AQFT was established by Brunetti, Fredenhagen and Verch \cite{BFV}, who acknowledged discussions with Hollands, Kay and Wald. These developments provided a common mathematical framework for formulating quantum field theories on arbitrary globally hyperbolic spacetimes and for controlling renormalization ambiguities.

The locally covariant formulation assigns an algebra \(\mathcal A(M)\) to every oriented, time-oriented globally hyperbolic spacetime \(M\). An orientation-preserving and time-orientation-preserving isometric embedding \(\psi:M\to N\) with causally convex image induces an injective unit-preserving star homomorphism. Consistency requires
\begin{equation}\label{eq:30}
\mathcal A(\psi\circ\chi)=\mathcal A(\psi)\circ\mathcal A(\chi),\qquad
\mathcal A(\mathrm{id}_M)=\mathrm{id}_{\mathcal A(M)}.
\end{equation}

In the language of category theory this is a covariant functor. Physically, the theory makes the same local assignment when a spacetime region is viewed inside a larger spacetime. Causal embeddings preserve the relevant causal relations. Einstein causality is imposed for spacelike separated images, and the time-slice axiom requires the map associated with a Cauchy embedding to be an isomorphism.

Restricting this assignment to subregions of one spacetime recovers a net. A state on \(N\) pulls back to a state on \(M\) by \(\omega_N\circ\mathcal A(\psi)\). This does not require a distinguished state to be assigned naturally on every background. A stationary spacetime may admit useful ground or thermal states, while a general background calls for other physically motivated choices.

The free scalar construction survives with \(P_{KG}=\Box_g+m^2+\xi R\), where \(R\) is the scalar curvature and \(\xi\) is the curvature coupling. Global hyperbolicity guarantees the advanced and retarded Green operators needed for the commutation relations. Backgrounds with timelike boundaries or failures of global hyperbolicity need additional boundary data and a modified formulation; they are outside the category just defined.

\subsection{The Hadamard condition}

An arbitrary positive state need not have physically
acceptable short-distance behaviour. For the free scalar
field on a curved spacetime, the Hadamard condition
characterises a class of states whose ultraviolet
singularities have the same local structure as those of
the Minkowski vacuum. It can be formulated either in
terms of the local singularity expansion of the two-point
function or, equivalently, by a condition on its
wavefront set \cite{FullingSweenyWald1978,FullingNarcowichWald1981,KayWald1991,Rad}.

Let $W$ denote the 2-point function of a state $\omega$ for the free scalar field. The distribution $W$ should be a bisolution for the equations of motion operator $P_{KG}$ and satisfy
\[
W(x,y)-W(y,x)=iE(x,y),
\]
where $E$ is the commutator function introduced in
Section~\ref{sec:commutators}. We write
$W=\frac{i}{2}E+\mu$,
as in Section~\ref{sec:Gaussian}, where $\mu$ is a real symmetric
bidistribution. Positivity of the state implies
that $W$ is of positive type.

The local Hadamard condition specifies the singular
behaviour of $W$. For the scalar field in four spacetime dimensions,
for every point there exists a geodesically convex
neighbourhood $U$ such that, for every integer
$N\geq 0$, the restriction of $W$ to
$U\times U$ can be written as
\begin{equation}\label{eq:hadamard-form}
	W(x,y)=
	\lim_{\varepsilon\downarrow0}
	\frac{1}{8\pi^2}
	\left[
	\frac{u(x,y)}{s_\varepsilon(x,y)}
	+
	\sum_{n=0}^{N}
	v_n(x,y)s(x,y)^n
	\log\left(\frac{s_\varepsilon(x,y)}{\ell^2}\right)
	\right]
	+w_N(x,y).
\end{equation}
Here $\sigma(x,y)$ denotes Synge's world function,
equal to one half of the signed squared geodesic
distance in our metric convention $(+---)$, and
\[
s(x,y)=-\sigma(x,y),\qquad
s_\varepsilon(x,y)=
s(x,y)+i\varepsilon\bigl(T(x)-T(y)\bigr)
+\varepsilon^2,
\]
where $T$ is a smooth time function increasing
towards the future. The limit is understood in
the sense of distributions.

The coefficients $u$ and $v_n$ are smooth, symmetric
biscalars determined locally by the geometry and
the Klein--Gordon operator through transport
equations, with $u(x,x)=1$. The parameter $\ell$
is an arbitrary reference length. The remainder
$w_N$ is of class $C^{2N+1}$ and contains
information depending on the state. Thus the
universal singular part is determined locally,
while the choice of state enters through the
remainder. In particular, the difference between
the two-point functions of any two Hadamard states
is smooth.

An equivalent characterisation was established by
Radzikowski \cite{Rad}. The Hadamard condition
is equivalent to the following restriction on the
wavefront set of the two-point function:
\begin{equation}\label{eq:hadamard-wf}
	\operatorname{WF}(W)
	=
	\left\{
	(x,k_x;y,-k_y)\in
	(T^*M\setminus\{0\})^2:
	(x,k_x)\sim(y,k_y),\
	k_x\in\overline V_+
	\right\}.
\end{equation}
Here $(x,k_x)\sim(y,k_y)$ means that $x$ and $y$
are connected by a null geodesic and that
$k_x$ and $k_y$ are nonzero, cotangent to it
and related by parallel transport.
The cone $\overline V_+$ denotes the closed
future light cone in cotangent space.

This microlocal spectrum condition provides a
local, generally covariant replacement for the
positive-frequency condition characterising the
Minkowski vacuum. It is particularly useful in
the construction of Wick polynomials and
time-ordered products, where the wavefront-set
restrictions control products of distributions.

In \cite{BFK96}, Brunetti, Fredenhagen and K\"ohler used the Hadamard condition and microlocal analysis in the construction of Wick polynomials.
These methods were subsequently extended to perturbative
renormalisation in \cite{BF0}. The later works of Hollands and
Wald \cite{HW,HW01,HW02} built on these results, incorporating
the principle of local covariance into the construction and
classification of Wick polynomials and time-ordered products.

The Hadamard condition is essential for defining local
observables such as the stress-energy tensor. After subtracting
the universal singular part, coincidence limits define Wick
powers and differentiated quadratic fields. 
The subtraction does not eliminate all ambiguities: local covariance,
scaling and other renormalisation requirements leave a controlled
freedom, such as adding multiples of $m^2$ and $R$ to the scalar
Wick square in four dimensions \cite{HW}.

For the stress tensor the corresponding ambiguities involve local curvature tensors and the parameters of the theory. Conservation must be imposed consistently. The resulting expectation value is the quantity used in semiclassical gravity. The algebraic formulation therefore connects a condition on state singularities to a finite local observable with explicitly identified renormalisation freedom.

\subsection{The response to geometry}

The time-slice axiom also permits comparison of dynamics before and after a compactly supported change of the metric. Passing through common past and future Cauchy regions defines an automorphism called relative Cauchy evolution. Its infinitesimal response is generated by the stress tensor, with the precise commutator sign set by the convention for that evolution \cite{BFV}. This identifies stress energy through the response of observables to geometry, complementing its definition from a Lagrangian. The background metric remains external in this construction; local covariance alone does not quantise gravity.

\subsection{Same physics in all spacetimes}
The principle of same physics in all spacetimes (SPASs) expresses the expectation that a quantum field theory should have the same physical content when formulated on different globally hyperbolic spacetimes. Although the functorial framework of locally covariant QFT \cite{BFV} provides a natural setting for this requirement, Fewster \cite{Fewster2012} and Fewster and Verch \cite{FewsterVerch2012} showed that local covariance alone is insufficient to guarantee it. They formulated the SPASs property, requiring that if two theories within a specified class are related by a natural embedding that is an isomorphism on one spacetime, then it must be an isomorphism on every spacetime. They further introduced dynamical locality, requiring agreement between the kinematic local algebras defined by spacetime embeddings and the dynamical local algebras determined by relative Cauchy evolution. They proved that the class of dynamically local theories satisfies the SPASs property, providing a stronger criterion for the physical consistency of locally covariant theories.

\section{Entanglement entropy in AQFT}

The algebraic approach to quantum field theory provides a
mathematically rigorous framework for studying entanglement
entropy. Since local von Neumann algebras in relativistic QFT
are typically factors of type $\mathrm{III}_1$, the ordinary
von Neumann entropy of a sharply localized region is not
intrinsically defined. Nevertheless, meaningful measures of
entanglement can be constructed using Araki's relative entropy,
mutual information, and the split property (see e.g.
\cite{HollandsSanders2018}). For two separated regions
$O_1,O_2$, the split property (see Section~\ref{sec:split}) permits the introduction of
intermediate type I factors, whose entropy can be studied
under suitable additional regularity assumptions.
Longo and Xu obtained explicit expressions for mutual
information in chiral conformal field theories and
established finiteness results for the entropy associated
with canonical intermediate type I factors
\cite{LongoXu2018,LongoXu2021}.
Subsequent work established asymptotic entropy estimates
and related singular limits of relative entropy to
subfactor indices and superselection structures
\cite{TangXu2022,BenedettiEtAl2024,Xu2025}.
These results demonstrate that entanglement encodes not only
local correlations but also global structural properties
of quantum field theories.

More recent developments concern rigorous entropy bounds
and their relation to energy conditions. Longo proved a
model-independent Bekenstein-type inequality for states
localized in a spacetime region of width $2R$,
\begin{equation}
	S\bigl(\omega_\Psi\|\omega_0\bigr)
	\leq 2\pi R\,E_\Psi,
\end{equation}
where the relative entropy is evaluated on the local
algebra, $\omega_0$ is the vacuum state, and $E_\Psi$
is the energy of the localized excitation
\cite{Longo2025Bekenstein}. Hollands and Longo extended
this inequality to approximately localized states and
charged representations. In the latter case, an additional
contribution involving the logarithm of the statistical
dimension of the superselection sector appears
\cite{HollandsLongo2026Bekenstein}.
Longo and Morinelli also derived bounds on local entropy
determined by spacetime symmetries
\cite{LongoMorinelli2026}, while Hollands, Longo and
Morsella obtained corresponding results for Klein--Gordon
wave packets \cite{HollandsLongoMorsella2026}.
In another development, Hollands and Longo derived an
explicit expression for the shape derivative of relative
entropy under deformations of an entangling surface,
leading to a new proof of the quantum null energy
condition (QNEC) \cite{HollandsLongo2025QNEC}.
Related inequalities constraining modular time evolution
were established by Hollands using half-sided modular
inclusions \cite{Hollands2026Modular}.

A different approach to the entropy problem has emerged
from connections between operator algebras and quantum
gravity. Witten proposed that gravitational constraints
and additional observer or energy degrees of freedom can
lead to a change in the type of the relevant observable
algebra. In the context of large-$N$ holography,
the inclusion of suitable subleading corrections leads
from an emergent type $\mathrm{III}_1$ algebra
$\mathcal M$ to a type $\mathrm{II}_\infty$ algebra,
described by the modular crossed product
\begin{equation}
	\mathcal N
	= \mathcal M\rtimes_{\sigma^\omega}\mathbb R,
\end{equation}
where $\sigma^\omega$ denotes the modular automorphism
group associated with a faithful normal state $\omega$
\cite{Witten2022Crossed}. Unlike a type III factor,
the crossed-product algebra possesses a faithful normal
semifinite trace, allowing a notion of entropy defined
up to a state-independent additive constant.
The mathematical properties of continuous entropy for
semifinite von Neumann algebras, including its relation
to relative entropy and subfactor indices, were further
investigated by Longo and Witten
\cite{LongoWitten2023Continuous}.

In a related construction, Chandrasekaran, Longo,
Penington and Witten showed that gravitationally dressed
observables associated with an observer in a static patch
of de Sitter space form a type $\mathrm{II}_1$ algebra.
The entropy of semiclassical states agrees, up to a
state-independent additive constant, with the generalized
entropy
\begin{equation}
	S_{\mathrm{gen}}
	= \frac{A}{4G_N}+S_{\mathrm{out}},
\end{equation}
and empty de Sitter space corresponds to a
maximum-entropy state
\cite{ChandrasekaranLongoPeningtonWitten2023}.
Witten subsequently proposed an observer-worldline
algebra as a candidate for a background-independent
algebra in quantum gravity, with entropy defined relative
to the Hartle--Hawking no-boundary state
\cite{Witten2024Background}.
Related ideas have been applied directly to QFT:
Kudler-Flam and collaborators developed a covariant
regularization of entanglement entropy based on modular
crossed products, establishing finite entropy differences
and deriving the Bekenstein bound and QNEC in this
framework \cite{KudlerFlamEtAl2025}. 

It is important to distinguish these constructions
from the original local algebras of AQFT: the type II
structure arises through an enlargement or modification
of the observable algebra, rather than a change in
the type of the original type III local algebra. A proposal to understand this enlargement in terms of \emph{quantum reference frames} has been formulated in \cite{FewsterJanssenLoveridgeRejznerWaldron2025} (see more in Section~\ref{sec:FV}).
\section{Interactions}\label{sec:9}

\subsection{Classical observables as functionals}

The structural results above do not require a preferred Lagrangian. To construct interacting examples perturbatively, however, it is useful to start from one. Perturbative AQFT combines local algebras with deformation quantisation and causal renormalisation \cite{BDF,Vienna,Book,Due19}. Its objects are formal power series rather than, automatically, completed \(C^*\)-algebras of bounded operators.

Begin with the space of smooth classical field configurations \(\varphi\). At first these are off shell: the field equation has not been imposed. An observable is a functional \(F[\varphi]\). A local functional has the form
\begin{equation}\label{eq:32}
F[\varphi]=\int f(x)\,\mathcal L(\varphi(x),\nabla\varphi(x),\ldots)\,d\mathrm{vol}_g(x),
\qquad f\in C_c^\infty(M).
\end{equation}

Only finitely many derivatives enter the local density. Functional derivatives encode how the observable changes under variations of the configuration. For the free scalar theory, the causal propagator defines the Peierls bracket, in our sign convention, by
\begin{equation}\label{eq:33}
\{F,G\}[\varphi]=\big\langle F^{(1)}[\varphi],E G^{(1)}[\varphi]\big\rangle.
\end{equation}

The pairing includes spacetime integration. This is a covariant way to express the classical Poisson bracket using causal response. For linear observables \(F_f[\varphi]=\int f\varphi\), it gives \(\{F_f,F_g\}=E(f,g)\). The field equation can be imposed afterwards by passing to the appropriate quotient. Retaining the off-shell description during the construction is particularly useful for renormalisation identities and gauge symmetry.

\subsection{Deforming the product}

The quantisation of free fields in pAQFT is formulated as
a deformation quantisation of the Poisson algebra of
classical observables. An early application of star-product
quantisation to the free scalar field was given by
Dito~\cite{Dito1990}, while its incorporation into the
perturbative algebraic framework was developed by
D\"utsch and Fredenhagen~\cite{DuetschFredenhagen2001}.
The extension of this construction to curved spacetimes
relies on the microlocal characterisation of Hadamard
two-point functions~\cite{Rad} and the
construction of Wick-polynomial algebras by Brunetti,
Fredenhagen and K\"ohler~\cite{BFK96}, further developed
by Brunetti and Fredenhagen~\cite{BF0} and Hollands and
Wald~\cite{HW}. In the functional formulation
of pAQFT, the quantum product is obtained by contracting
functional derivatives with a Hadamard bidistribution,
and different choices lead to isomorphic star-product
algebras. Systematic accounts are given in
\cite{FredenhagenRejzner2015,Book,Due19}.

Choose a Hadamard two-point function \(W\) for the free field, normalised without a factor of \(\hbar\), so that \(W-W^{\mathrm T}=iE\). For regular (i.e. all their functional derivatives are smooth compactly supported kernels) polynomial functionals, define
\begin{equation}\label{eq:34}
	(F\star_W G)[\varphi]
	=\sum_{n=0}^{\infty}\frac{\hbar^n}{n!}
	\big\langle F^{(n)}[\varphi],W^{\otimes n}G^{(n)}[\varphi]\big\rangle.
\end{equation}

This is the functional version of contracting fields in Wick's theorem. The zeroth term is ordinary multiplication. Antisymmetrising its first correction gives
\begin{equation}\label{eq:35}
	\frac{F\star_WG-G\star_WF}{i\hbar}
	=\{F,G\}+O(\hbar).
\end{equation}

For linear functionals the higher derivatives vanish, so the commutator is exactly \(i\hbar E(f,g)\). Thus the canonical commutation relations reappear as the first and, in this case, only quantum correction.

Local polynomial functionals have distributional derivatives. Microcausal wavefront-set restrictions make contractions well defined at each configuration, but need not ensure smooth field dependence: even the Peierls bracket of regular functionals can fail to be smooth. \cite{HawkinsRejznerVisser2026} introduce \emph{equicausal functionals}, whose derivative families are equicontinuous, as linear functionals on appropriate distribution spaces, over compact sets of configurations. This controls the variation of singularities with the field. This class contains local functionals and Wick polynomials, closes under the Peierls bracket and formal \(\star_W\)-product, and supports a chain-homotopy proof of the time-slice axiom in the free-field setting considered.

Changing \(W\) to another Hadamard two-point function changes it by a smooth symmetric kernel. The resulting star products are related by an explicit normal-ordering isomorphism. The construction therefore does not require one particular vacuum. 

\subsection{Time ordering and local renormalisation}
Introducing  interactions requires time-ordered products in addition to  the star product. Time-ordered products of functionals with disjoint supports are well-defined. Singularities arise when supports coincide. This can be reduced to the problem of defining certain quantities as distributions on $M^n$ ($n\in \NN$ characterises the number of interaction vertices)  where they are a priori defined only outside diagonals. Causal perturbation theory treats renormalisation as extension of distributions to these diagonals \cite{BF0,EG}.

For a compactly supported interaction $V$, for example
\begin{equation}\label{eq:36}
V[\varphi]=\frac{\lambda}{4!}\int g(x)\varphi(x)^4\,d^4x,
\qquad g\in C_c^\infty(M),
\end{equation}

the formal local S-matrix is
\begin{equation}\label{eq:37}
S(V)=\sum_{n=0}^{\infty}\frac1{n!}
\left(\frac{i}{\hbar}\right)^nT_n(V^{\otimes n}),
\qquad T_0=\mathbf1.
\end{equation}

The S-matrix is formal in the coupling, and the coefficients $T_n$ are called \emph{$n$-fold time-ordered products}.
The S-matrix satisfies the  causal factorisation law:
\begin{equation}\label{eq:38}
S(F+G+H)=S(F+G)\star S(G)^{-1}\star S(G+H)
\end{equation}
if the support of \(F\) is \emph{not earlier} than the support of \(H\); equivalently, \(\operatorname{supp}F\) does not meet the causal past of \(\operatorname{supp}H\). The inverse and the products on the right are with respect to \(\star\). This identity recursively determines time-ordered products away from coinciding configurations. Extending to coincidence introduces a finite-dimensional ambiguity at each fixed order and scaling degree.

\begin{exa}
 In four dimensions a scalar Feynman propagator has short-distance scaling degree two. One subdistribution appearing at second order in a quartic interaction is its square, initially defined away from the origin in the relative coordinate. It has scaling degree four. An extension preserving that scaling degree can therefore change only by
\begin{equation}\label{eq:39}
t_{\mathrm{ren}}(x)\longmapsto t_{\mathrm{ren}}(x)+c\,\delta^{(4)}(x).
\end{equation}

No derivative of the delta distribution is allowed here because it would raise the scaling degree. On insertion into the graph, the delta distribution brings the two vertices together and produces a local counterterm. Symmetry, reality and other normalisation requirements further constrain the coefficient. This example explains why an ultraviolet ambiguity can be represented by a local change of parameters. More singular graphs allow finitely many derivatives of delta distributions, with their orders bounded by power counting.
\end{exa}

The method produces finite distributions at each perturbative order. It does not require assigning a physical meaning to divergent intermediate quantities, and it exposes exactly where renormalisation choices enter. Different admissible prescriptions are related by local redefinitions of the interaction and fields. This is the setting for the perturbative renormalisation group. Note that having finitely many parameters at each order does not mean that the ambiguity can be reduced to finitely many parameters globally, after one sums all orders. This is the case only for \emph{renormalizable} theories.

\subsection{Recovering interacting local observables}

The local S-matrix is a device for constructing observables even when no global scattering matrix is available. Define the relative S-matrix and the interacting (with interaction $V$) observable associated with a perturbation \(F\) by
\begin{equation}\label{eq:40}
S_V(F)=S(V)^{-1}\star S(V+F),\qquad
R_V(F)=\left.\frac{\hbar}{i}\frac{d}{dt}S_V(tF)\right|_{t=0}.
\end{equation}

Relative S-matrices with perturbations supported in \(O\) generate the interacting local algebra. Causal factorisation controls the comparison of different compactly supported switching functions that agree where the local construction requires them to agree. The resulting algebras can be identified through compatible isomorphisms. This gives an algebraic adiabatic construction of the local theory \cite{Vienna}.

This local construction must be distinguished from the existence of a state or a scattering operator after sending a coupling to a constant over all spacetime. Such global limits face additional infrared and state-dependent questions. Likewise, well-defined formal coefficients do not prove convergence of the perturbation series at a nonzero numerical coupling.

\subsection{Dynamical \texorpdfstring{$C^*$}{C*}-algebras}\label{sec:dynamical-cstar}

Buchholz and Fredenhagen use the structural identities of local S-matrices to define nonperturbative \emph{dynamical \(C^*\)-algebras}~\cite{BuchholzFredenhagen2020}. Consider a real scalar field on Minkowski space with a standard kinetic term and a polynomial interaction potential. Associate a unitary symbol \(S_L(F)\) with each compactly supported real local polynomial functional \(F\) without field derivatives, interpreted as a local perturbation of the dynamics specified by the Lagrangian \(L\). Set \(\hbar=1\) in this subsection.

For a compactly supported field shift \(\psi\), write \(F^\psi[\varphi]=F[\varphi+\psi]\) and define the relative action
\begin{equation}\label{eq:dynamical-relative-action}
	\delta L(\psi)[\varphi]=L(\chi)[\varphi+\psi]-L(\chi)[\varphi],
	\qquad L(\chi)[\varphi]=\int_M\chi\,\mathcal L[\varphi]\,d^dx.
\end{equation}
Here \(\chi\in C_c^\infty(M)\) equals one near \(\operatorname{supp}\psi\); the difference is independent of its choice. The defining relations are causal factorisation,
\begin{equation}\label{eq:dynamical-causality}
	S_L(F+G+H)=S_L(F+G)S_L(G)^{-1}S_L(G+H),
\end{equation}
when a Cauchy surface separates \(\operatorname{supp}F\) to its future from \(\operatorname{supp}H\) to its past, and the dynamical relation
\begin{equation}\label{eq:dynamical-schwinger-dyson}
	S_L(F^\psi+\delta L(\psi))
	=S_L(F)S_L(\delta L(\psi))
	=S_L(\delta L(\psi))S_L(F).
\end{equation}
The latter is an integrated Schwinger--Dyson identity. The elements \(S_L(\delta L(\psi))\) are central; imposing \(S_L(\delta L(\psi))=\mathbf1\) gives the source-free on-shell quotient. This formulation and its extension to fermionic theories are discussed in~\cite{BrunettiDuetschFredenhagenRejzner2022}.

Fixing \(S_L(c)=e^{ic}\mathbf1\) for constant functionals and completing the resulting group \(*\)-algebra in its maximal \(C^*\)-norm gives \(\mathcal A_L\). Its local subalgebras are
\begin{equation}\label{eq:dynamical-local-net}
	\mathcal A_L(O)=C^*\!\bigl(\{S_L(F):\operatorname{supp}F\subset O\}\bigr).
\end{equation}
They satisfy isotony and Einstein causality, with Poincar\'{e} covariance when the Lagrangian is invariant. Spacelike commutativity follows by applying causal factorisation in both orders. For the free Lagrangian, suitably phased linear generators obey the Weyl relations. For a compactly supported interaction \(V\), the elements \(S_L(V)^{-1}S_L(V+F)\) implement a local interaction picture, connecting this construction with~\eqref{eq:40}~\cite{BuchholzFredenhagen2020}.

The construction yields genuine \(C^*\)-algebras at fixed coupling. It does not itself establish the existence of physically admissible vacuum or thermal states for general interactions; this becomes a problem in their representation theory. Extensions accommodate kinetic perturbations, with the causal relations adjusted to the perturbed metric and Fock representations constructed for admissible quadratic perturbations~\cite{BuchholzFredenhagen2021}, and Fermi fields through a functorial use of auxiliary Grassmann variables~\cite{BrunettiDuetschFredenhagenRejzner2022}. Global symmetries were treated in \cite{BrunettiDuetschFredenhagenRejzner2023,BrunettiDuetschFredenhagenRejzner2024}.

\subsection{Gauge theories and the BV formalism}

The treatment of local gauge theories within perturbative algebraic
quantum field theory (pAQFT) combines causal perturbation theory
with the BRST and Batalin--Vilkovisky (BV) formalisms.
An early systematic construction was given by Hollands
\cite{Hollands2008YangMills}, who established perturbative
renormalisation of Yang--Mills theory on globally hyperbolic curved
spacetimes, using Ward identities and BRST cohomology to
characterise gauge-invariant observables.
Fredenhagen and Rejzner subsequently developed a locally covariant
formulation of the classical BV complex and its perturbative
quantisation \cite{FredenhagenRejzner2012,FredenhagenRejzner2013},
identifying physical observables through cohomological methods
and incorporating gauge symmetries into the algebraic framework.
In this approach, the renormalised quantum master equation and
the anomalous Master Ward identity provide the appropriate
framework for analysing gauge invariance and anomalies, while
the quantum BV differential is related, on shell, to the
interacting BRST charge \cite{Rejzner2015LocalSymmetry} (further improved by \cite{RejznerSchiavina2026}).
These methods have been applied to perturbative quantum gravity
\cite{BrunettiFredenhagenRejzner2016} and further developed
through the study of anomaly consistency conditions and their
associated $L_\infty$-algebraic structures
\cite{Froeb2019Anomalies,BrunettiDuetschFredenhagenRejzner2024}.
More recently, the framework has been extended towards
BV--BFV quantisation in the presence of boundaries
\cite{RejznerSchiavina2026}.

For a gauge theory the configuration space contains redundant descriptions. A direct quotient by gauge transformations can be singular and can obscure locality. The Batalin--Vilkovisky, or BV, formalism instead enlarges the space with ghosts and antifields, resulting in a graded algebra $\mathcal B=\bigoplus_k\mathcal B^k$,
 and introduces a differential \(s\) of degree one. Its square vanishes. Under the usual regularity assumptions, degree-zero cohomology describes gauge-invariant on-shell observables:
\begin{equation}\label{eq:41}
H^0(s)=\frac{\ker(s:\mathcal B^0\to\mathcal B^1)}
{\operatorname{im}(s:\mathcal B^{-1}\to\mathcal B^0)}.
\end{equation}

In typical physics applications, the degree-zero cohomology describes gauge-invariant observables on the space of classical solutions, modulo gauge transformations. At the classical level the master equation for the extended action guarantees the consistency of the differential. At the quantum level renormalised Ward identities and the quantum master equation express compatibility of gauge symmetry with the interacting product. For more detailed discussion, see e.g. \cite{Rej20}.

An anomaly is an obstruction to imposing these identities consistently, and its cohomology class determines whether a change of local counterterms can remove it. Gauge fixing supplies a workable hyperbolic free operator, while the cohomological construction tracks the intended independence of physical observables from that choice. These are the main reasons homological methods enter perturbative AQFT. 

\section{Measurement in AQFT}\label{sec:FV}
\subsection{Measurement}
\begin{figure}[htbp]
	\centering
\begin{tikzpicture}[font=\small,>=Latex,line cap=round]
	\coordinate (bot) at (0,-2.6);
	\coordinate (left) at (-2.7,0);
	\coordinate (top) at (0,2.6);
	\coordinate (right) at (2.7,0);
	\fill[before!20] (bot)--(left)--(0,-0.72)--(right)--cycle;
	\fill[after!20] (top)--(left)--(0,0.72)--(right)--cycle;
	\fill[gray!10] (left)--(0,-0.72)--(right)--(0,0.72)--cycle;
	\draw[gray!85,thick] (bot)--(left)--(top)--(right)--cycle;
	\draw[gray!50,dashed] (left)--(0,0.72)--(right);
	\draw[gray!50,dashed] (left)--(0,-0.72)--(right);
	\fill[interaction!65,draw=interaction!80!black,thick] (0,0) ellipse (0.64 and 0.25);
	\node[font=\bfseries] at (0,0) {$K$};
	\node[before!70!black,align=center] at (0,-1.72) {pre-coupling\\region};
	\node[after!70!black,align=center] at (0,1.3) {post-coupling\\region};
	\draw[->,thick] (-3,-2.25)--(-3,2.25) node[above] {$t$};
	\node[below,gray!70!black] at (0,-2.78) {spacetime $M$};
	\node[draw=before!75!black, fill=before!7, rounded corners=2pt,
	inner sep=8pt,minimum width=5.2cm,align=center] (pre) at (6,-1.73)
	{Prepare the state $\omega\otimes\sigma$ early};
	\node[draw=interaction!90!black,fill=interaction!15,rounded corners=2pt,
	inner sep=8pt,minimum width=5.2cm,align=center] (coupling) at (6,0)
	{interaction confined to $K$\\coupled theory $\mathfrak C$};
	\node[draw=after!75!black,fill=after!7,rounded corners=2pt,
	inner sep=8pt,minimum width=5.2cm,align=center] (readout) at (6,1.73)
	{Measure $B$ late};
	\draw[->,thick,gray!80!black] (pre)--node[right,xshift=3pt] {coupling} (coupling);
	\draw[->,thick,gray!80!black] (coupling)--node[right,xshift=3pt] {readout} (readout);
\end{tikzpicture}
	\caption{Schematic of the Fewster--Verch measurement scheme.
		A system field $\mathfrak A$ and a probe field $\mathfrak B$ interact
		only within the compact coupling region $K$. The subsequent probe
		readout determines an induced system observable
		$\varepsilon_\sigma(B)$ via the scattering morphism $\Theta$.}
	\label{fig:fewster-verch-measurement}
\end{figure}
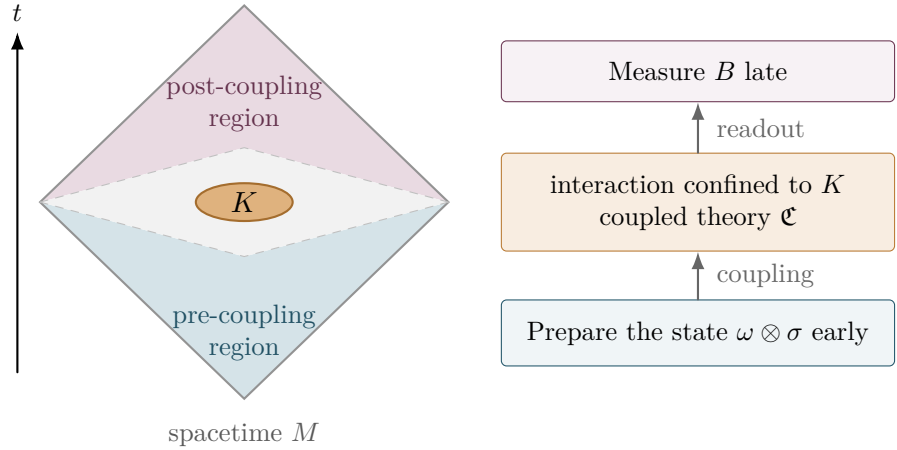

The measurement framework of Fewster and Verch
\cite{FewsterVerch2020} provides a local and covariant account of quantum
measurements in algebraic quantum field theory, in which the measuring
apparatus is modelled by a second quantum field theory rather than an
external projection postulate. Let $\mathfrak A$ describe the system and
$\mathfrak B$ the probe, with uncoupled theory
$\mathfrak U=\mathfrak A\otimes\mathfrak B$. The two theories are
coupled in a compact spacetime region $K\subset M$, resulting in a
coupled theory $\mathfrak C$ that agrees with $\mathfrak U$ in the
pre- and post-coupling regions. The time-slice axiom gives
identifications $\tau^\pm:\mathfrak U(M)\longrightarrow\mathfrak C(M)$,
which define the scattering morphism
\begin{equation}
	\Theta=(\tau^-)^{-1}\circ\tau^+.
\end{equation}
The probe $\Bf$ is prepared in some initial state $\sigma$ uncorrelated to an $\Af$ state $\omega$ in the \emph{pre-coupling region} (see Figure~\ref{fig:fewster-verch-measurement}). It induces a state $((\tau^{-})^{-1})^*(\omega\otimes \sigma)$ on $\mathfrak{C}(M)$.
A probe observable $B$ is measured in the \emph{post-coupling region}. It can be identified with an observable of the coupled theory by means of $\tau^+(1\otimes B)\in \mathfrak{C}(M)$.

Putting things together, we obtain
\[
((\tau^{-})^{-1})^*(\omega\otimes \sigma)(\tau^+(1\otimes B))=\omega\otimes\sigma(\Theta (1\otimes B))\,,
\]
and interpret it as the expected value of the \emph{induced system observable} $\varepsilon_\sigma(B)
=\eta_\sigma\bigl(\Theta(\mathbf 1\otimes B)\bigr)$ in the state $\omega$. Here $\eta_\sigma(A\otimes B):=\sigma(B)A$.

The map $B\mapsto \varepsilon_\sigma(B)$ is unital and completely positive and the induced observable can be localized in a region containing $K$. The procedure described above is \emph{a measurement scheme for the system observable $ \varepsilon_\sigma(B)$}.

The theory also supplies instruments describing selective and non-selective state updates. Its causal factorisation
property ensures that measurement schemes in causally ordered regions
compose consistently; schemes with causally disjoint coupling regions
may be composed in either order. Thus measurement statistics and state
updates are obtained from a local dynamical interaction compatible with
relativistic causality, without positing an instantaneous physical
collapse.

\subsection{Quantum reference frames}
Quantum reference frames (QRFs) have been studied from several
complementary perspectives, including the quantum-information
approach to reference-frame asymmetry and superselection rules
\cite{BRS07}, symmetry-invariant relational observables
\cite{LMB18}, and transformations between quantum perspectives,
formulated either directly \cite{GCB19,dHG20} or within a
perspective-neutral framework based on constrained systems
\cite{VHGC20}. The approach discussed below builds on the
operational formulation in terms of covariant POVMs and
invariant observables \cite{LMB18,CaretteGlowackiLoveridge2025}, which is particularly
well suited to the operator-algebraic setting of AQFT.

On a spacetime with a group $G$ of isometries, a measurement scheme
whose interaction is localised in $K$ is carried by the group action to
a scheme localised in $gK$. Even if the field theories are covariant,
choosing $K$ singles out a position relative to the background. The
operational quantum-reference-frame approach
\cite{CaretteGlowackiLoveridge2025,FewsterJanssenLoveridgeRejznerWaldron2025}
replaces this external choice with a physical quantum system. A frame
is described by a Hilbert space $\mathcal H_R$, a strongly continuous
unitary representation $U_R$ of $G$, and a normalised POVM
$\mathsf{E}:\operatorname{Bor}(\Sigma)\to B(\mathcal H_R)$, for a homogeneous
$G$-space $\Sigma$, satisfying
\begin{equation}
	U_R(g)\mathsf{E}(X)U_R(g)^*=\mathsf{E}(gX).
\end{equation}
Measurements of the field are then specified relative to the quantum
frame, and their physical representatives belong to the algebra of
joint $G$-invariant observables, schematically
$(\mathfrak A_S\otimes\mathfrak A_R)^G$. This relativisation
procedure leads naturally to observables that need not be localised in
an arbitrarily prescribed spacetime region. It can also change the
operator-algebraic structure: under suitable thermal (KMS)
assumptions, the invariant algebra admits a semifinite trace, which
becomes finite under additional compatible thermal assumptions on the
reference frame \cite{FewsterJanssenLoveridgeRejznerWaldron2025}. 

QRFs can also be constructed from degrees of freedom associated with quantum fields. A recent example is provided by Fewster, Janssen and Rejzner \cite{FJR26}, who develop an algebraic approach to electromagnetism on finite Cauchy lenses, a class of spacetimes with boundaries and corners. In this setting, the relevant symmetry is the group of large gauge transformations rather than spacetime isometries. Auxiliary boundary degrees of freedom serve as quantum reference frames for these transformations, allowing the construction of gauge-invariant semi-local observables through a relativisation map at the level of \(C^*\)-algebras. These observables involve both bulk and boundary degrees of freedom and therefore extend the usual notion of localisation in AQFT.

These developments show how QRFs could provide a relational description of localisation and measurement in AQFT, replacing externally prescribed localisation regions with localisation relative to a physical quantum system. They also reveal a connection between operational measurement theory and the operator-algebraic structure of quantum field theory, suggesting new approaches to understanding local observables, modular theory and entanglement entropy.

\vspace{-3ex}

{
\bibliographystyle{amsalpha}
\bibliography{References}
}

\end{document}